\RequirePackage{ifpdf}
\ifpdf 
\documentclass[pdftex]{sigma}
\else
\documentclass{sigma}
\fi

\numberwithin{equation}{section}

\newtheorem{thm}{Theorem}[section]
\newtheorem{cor}[thm]{Corollary}
\newtheorem{lem}[thm]{Lemma}
\newtheorem{prop}[thm]{Proposition}
\newtheorem{conj}[thm]{Conjecture}
{\theoremstyle{definition}
\newtheorem{rem}[thm]{Remark}
\newtheorem{exam}[thm]{Example}
}

\begin{document}

\allowdisplaybreaks

\renewcommand{\thefootnote}{$\star$}

\renewcommand{\PaperNumber}{062}

\FirstPageHeading

\ShortArticleName{On Algebraically Integrable Dif\/ferential Operators
on an Elliptic Curve}

\ArticleName{On Algebraically Integrable Dif\/ferential Operators\\
on an Elliptic Curve\footnote{This paper is a
contribution to the Special Issue ``Relationship of Orthogonal Polynomials and Special Functions with Quantum Groups and Integrable Systems''. The
full collection is available at
\href{http://www.emis.de/journals/SIGMA/OPSF.html}{http://www.emis.de/journals/SIGMA/OPSF.html}}}

\Author{Pavel ETINGOF~$^\dag$ and Eric RAINS~$^\ddag$}

\AuthorNameForHeading{P.~Etingof and E.~Rains}

\Address{$^\dag$~Department of Mathematics, Massachusetts Institute of Technology,\\
\hphantom{$^\dag$}~Cambridge, MA 02139, USA}
\EmailD{\href{mailto:etingof@math.mit.edu}{etingof@math.mit.edu}}
\URLaddressD{\url{http://www-math.mit.edu/~etingof/}}

\Address{$^\ddag$~Department of Mathematics, California Institute of Technology, Pasadena, CA 91125, USA}
\EmailD{\href{mailto:rains@caltech.edu}{rains@caltech.edu}}

\ArticleDates{Received April 25, 2011, in f\/inal form June 30, 2011;  Published online July 07, 2011}

\Abstract{We study dif\/ferential operators on an elliptic curve of order
higher than 2 which are algebraically integrable (i.e., f\/inite gap).
We discuss classif\/ication of such operators of order 3 with one pole,
discovering exotic operators on special elliptic curves def\/ined over~${\mathbb Q}$
which do not deform to generic elliptic curves. We also study algebraically integrable operators of higher order
with several poles and with symmetries, and (conjecturally) relate them to crystallographic elliptic Calogero--Moser systems
(which is a generalization of the results of Airault, McKean, and Moser).}


\Classification{35J35; 70H06}

 \rightline{\it To Igor Moiseevich Krichever on his 60th birthday}

\renewcommand{\thefootnote}{\arabic{footnote}}
\setcounter{footnote}{0}

\section{Introduction}

In this paper we study dif\/ferential operators
\[
L=\partial^n+a_2(z)\partial^{n-2}+\cdots+a_n(z)
\] which are
algebraically integrable (i.e., there exists a nonzero dif\/ferential
operator $M$ of order relatively prime to $L$ such that
$[L,M]=0$). Such operators were f\/irst studied in \cite{BuC} and became
a focus of attention since the seventies, as they provide explicit
solutions to the Gel'fand--Dickey hierarchy (in particular, the KdV
hierarchy for $n=2$ and the Boussinesq hierarchy for $n=3$; see
\cite{DMN,DKMN} and references therein). A~general
classif\/ication of such operators was obtained in~\cite{Kr1}.

We are interested in making this classif\/ication more explicit
in the special case when the coef\/f\/icients $a_i(z)$ are meromorphic functions
on an elliptic curve $E$. For instance, in the simplest nontrivial case $n=2$
and a single pole, it is well known that the only algebraically integrable
operator, up to equivalence, is the Lam\'e operator
\[
L=\partial^2-m(m+1)\wp(z),
\]
where $m$ is an integer; its algebraic integrability was discovered by Hermite.
In the case $n=2$ and multiple poles, the answer
is much more interesting (see \cite[Section~4.1]{CEO}, as well as
\cite{GW,GUW} and references therein).

We study the problem of classif\/ication of algebraically integrable operators $L$
for $n>2$. It turns out that already in the case $n=3$ and a single pole,
the situation is much richer and more complicated than for $n=2$;
in particular, there exist algebraically integrable operators~$L$ of third order with one pole
def\/ined on an inf\/inite family of special elliptic curves over~$\Bbb Q$, which do not deform
to operators (with one pole) on a generic elliptic curve. We provide a list of third order
algebraically integrable operators with one pole which is conjecturally complete, and
 state some results and conjectures concerning operators with
several poles. In particular, we conjecture that in the special
case of operators with symmetries, algebraically integrable
operators are described in terms of the classical
crystallographic elliptic Calogero--Moser systems introduced in~\cite{EFMV}.

The paper is organized as follows. In Section~\ref{section2}, we give an
exposition of the general theory of algebraically integrable
operators, in particular those on an elliptic curve;
the results here are mostly well known, but for reader's
convenience we give
an exposition based on dif\/ferential Galois theory similar to one
in~\cite{CEO}. In this section we also propose a general conjecture on the
classif\/ication of operators with one pole. In Section~\ref{section3},
we present computational results for third order algebraically
integrable operators with one pole, and give a conjectural classif\/ication of such
operators. Finally, in Section~\ref{section4} we discuss operators with
several poles, and state a~conjecture on the connection with the
systems of~\cite{EFMV}.

\section{The general theory of algebraically integrable operators}\label{section2}

In this subsection we review the basics on algebraically integrable
ordinary dif\/ferential operators. Most of this material is well known; we refer the reader to \cite{Kr1, SW, Pr,
Tr, GW, CEO, Ch, GUW} and references therein.

\subsection{Def\/inition of algebraic integrability and meromorphicity of coef\/f\/icients}\label{section2.1}

Consider the dif\/ferential operator
\[
L=\partial^n+a_1(z)\partial^{n-1}+\cdots +a_n(z),
\]
where $a_i(z)$ are smooth functions on some interval in $\Bbb R$.
Note that the coef\/f\/icient $a_1$ can be gauged away
by conjugation of $L$ by $e^{\frac{1}{n}\int a_1(z)dz}$. So without loss of generality, we may
(and will) assume that $a_1=0$, i.e.
\begin{gather*}
L=\partial^n+a_2(z)\partial^{n-2}+\cdots+a_n(z).
\end{gather*}

Recall that $L$ is called {\it algebraically integrable}
(or algebro-geometric) if there exists a nonzero dif\/ferential operator $M$
of order relatively prime to $L$ such that $[L,M]=0$~\cite{BuC}.
Note that up to scaling $M$ is necessarily monic (i.e., has leading coef\/f\/icient~$1$).

For example, if $n=2$ then $L=\partial^2+u(z)$, and if $L$ is algebraically integrable then~$u$
is called a {\it finite-gap potential}.

\begin{thm}[\protect{\cite[Theorem 6.10]{SW}, see also \cite{Kr1}}]\label{regsin}
If $L$ is algebraically integrable,
then~$a_i(z)$ extend to meromorphic functions on the complex plane.
Moreover, the order of each pole of the function~$a_i$ in~$\Bbb C$
is at most~$i$, for $i=2,\dots ,n$; in other words,
the operator $L$ has regular $($or Fuchsian$)$ singularities in $\Bbb C$.
\end{thm}

\subsection[The indices of $L$]{The indices of $\boldsymbol{L}$}\label{section2.2}

Let $a_i(z)=b_iz^{-i}(1+O(z))$ near $z=0$.
Then by rescaling $z$ the operator $L$ can be degenerated into the operator
with rational coef\/f\/icients
\begin{gather}\label{ratope}
L_0=\partial^n+b_2z^{-2}\partial^{n-2}+\cdots +b_nz^{-n}=0.
\end{gather}

Consider the dif\/ferential equation
\[
L_0z^m=0.
\]
This equation is equivalent to the algebraic equation
\[
P_L(m)=0,
\]
where
\[
P_L(m)=m(m-1)\cdots (m-n+1)+b_2m(m-1)\cdots (m-n+3)+\dots +b_n.
\]
Let $m_j$, $j=0,\dots ,n-1$, be the roots
(with multiplicities) of the polynomial $P_L$, i.e.
\[
P_L(s)=(s-m_0)\cdots (s-m_{n-1}).
\]
The numbers $m_j$ are called the {\it indices} of $L$ at $0$.
They are arbitrary numbers satisfying the relation
\[
\sum_{j=0}^{n-1}m_j=n(n-1)/2.
\]
Obviously, the indices uniquely determine the coef\/f\/icients $b_i$.

Similarly, one def\/ines the indices of $L$ at any point $z_0$.

\begin{exam} The indices of $L$ at a regular point are $0,1,\dots ,n-1$.
\end{exam}

\subsection{Algebraic integrability of homogeneous rational operators}\label{section2.3}

\begin{prop}\label{integrat}
The operator $L_0$ given by formula \eqref{ratope} is algebraically integrable if and only if
the indices $m_j$ are integers which are distinct modulo $n$.
\end{prop}

\begin{proof}
It is easy to see (see e.g.~\cite{BEG}) that the operator $L_0$ is
algebraically integrable if and only if the equation $L_0\psi=\mu^n\psi$ admits
a Baker--Akhiezer solution of the form $F(\mu z)$, where
\[
F(x)=e^xQ(1/x),
\]
and $Q$ is a polynomial such that $Q(0)=1$.\footnote{E.g., if
$L_0$ is algebraically integrable, then it admits a homogeneous commuting
operator $M_0$ of relatively prime order
$m$ such that $L_0^m=M_0^n$, and the system of
dif\/ferential equations $L_0\psi=\mu^n\psi$, $M_0\psi=\mu^m\psi$
can be reduced to a f\/irst order scalar equation, which has a
solution of the required form by Euler's formula.}
Solving the equation by the power series method, we see that this happens
if and only if $m_j$ are distinct modulo~$n$ (and thus represent each residue class exactly once).
\end{proof}

\begin{cor}\label{intcon}
If $L$ is algebraically integrable then the indices $m_j(z_0)$ of $L$ at every point~$z_0$
are integers which are distinct modulo~$n$.
\end{cor}

\begin{proof} Suppose $L$ is algebraically integrable.
Since by Theorem~\ref{regsin},
the commuting opera\-tor~$M$ has regular singularities at $z_0$,
its rational degeneration $M_0$ is well def\/ined,
and commutes with~$L_0$, so~$L_0$ is algebraically integrable.
So the result follows from Proposition~\ref{integrat}.
\end{proof}

When $m_j$ are integers,
we will order them in the increasing order,
$m_0\le m_1\le\cdots \le m_{n-1}$.
It is also convenient to
introduce the ``gaps'' $q_j:=m_j-m_{j-1}$, $j=1,\dots ,n-1$, which clearly determine $m_j$.
In the integrable case, these are nonnegative integers not divisible by $n$.

\subsection{Algebraic integrability of dif\/ferential operators on elliptic curves}\label{section2.4}

Let $\Gamma\subset \Bbb C$ be a lattice, and $E=\Bbb C/\Gamma$ be the corresponding elliptic curve.
Assume that~$a_i(z)$ are rational functions on~$E$ (i.e., elliptic functions).

\begin{thm}\label{fourcond}
The following conditions on $L$ are equivalent.
\begin{enumerate}\itemsep=0pt
\item[$(i)$] $L$ is algebraically integrable.

\item[$(ii)$] The monodromy of the equation
\begin{gather}\label{eige}
L\psi=\lambda \psi
\end{gather}
around every pole of $L$
in $E$ is trivial for any eigenvalue $\lambda\in \Bbb C$.

\item[$(iii)$] The monodromy group of equation \eqref{eige}
is upper triangular in some basis.

\item[$(iv)$] For generic $\lambda$ $($i.e., outside of finitely many values$)$, equation \eqref{eige}
has a basis of solutions of the form
\begin{gather}\label{eige1}
\psi(z)=e^{\beta z}\prod_{i=1}^m \frac{\theta(z-\alpha_i)}{\theta(z-\beta_i)},
\end{gather}
where $\theta$ is the first Jacobi theta-function.
\end{enumerate}
\end{thm}

\begin{proof}
$(i)$ $\implies$ $(ii)$. Since $L$ is algebraically integrable,
by Corollary~\ref{intcon}, the indices of $L$ at every pole are integers.
Hence the monodromy matrices of equation~(\ref{eige})
around the poles of $L$ are unipotent. Also, it follows from~\cite{BEG} that
the dif\/ferential Galois group of~(\ref{eige}) for generic~$\lambda$
is an algebraic torus. Since monodromy matrices belong to the dif\/ferential Galois group,
and since every unipotent element of a torus is trivial, we conclude that the monodromy matrices of~(\ref{eige})
around the poles of $L$ are trivial for generic, hence for all $\lambda$.

$(ii)$ $\implies$ $(iii)$. If $(ii)$ holds, the monodromy group of (\ref{eige}) is Abelian,
hence is upper triangular in some basis.

$(iii)$ $\implies$ $(iv)$. Assume $(iii)$ holds. Since $L$ has regular
singularities, the dif\/ferential Galois group of equation (\ref{eige})
is the Zariski closure of the monodromy group. Hence the
dif\/ferential Galois group of equation (\ref{eige}) is triangular as well.
But it is shown in \cite{BEG} that the dif\/ferential Galois group of (\ref{eige}) is reductive for generic $\lambda$. Hence
this group is Abelian, and is a torus for generic $\lambda$. Thus there is a fundamental system of solutions
of the form (\ref{eige1}) (see \cite{BEG}).

$(iv)$ $\implies$ $(i)$. If $(iv)$ holds, then the monodromy of equation (\ref{eige}) around poles is trivial for generic $\lambda$.
Hence it is trivial for all $\lambda$, and the monodromy group, hence the dif\/ferential Galois group
of (\ref{eige}) is Abelian. So by \cite{BEG}, there
exists a nonzero dif\/ferential operator $M$ of order coprime to the order of $L$ such that $[L,M]=0$.
\end{proof}

\begin{rem} \quad{}

1. This theorem is similar to Theorem~5.9 in~\cite{Ch}, which goes back to~\cite{CEO}.

2. A similar theorem, with the same proof, holds in the trigonometric (nodal) and rational (cuspidal) case, i.e.,
when the coef\/f\/icients of~$L$ are rational functions on the nodal or cuspidal curve of arithmetic genus~$1$ which are
regular at inf\/inity. More precisely, in the rational case, since the singularity at inf\/inity is irregular,
we must add the Stokes matrix at inf\/inity to
the monodromy group, and we should also replace~$\theta(z)$ with~$\sin(z)$ and~$z$ in the
trigonometric and rational cases, respectively. We note that in the trigonometric and rational case,
the implication $(iv)$ $\implies$ $(i)$ of Theorem~\ref{fourcond} was proved in~\cite{W}.
\end{rem}

\subsection{Operators with one pole}\label{section2.5}

In this subsection we will consider the special case when $L$ has only one pole, at the point $0\in E$.

\subsubsection{Second order operators with one pole}\label{section2.5.1}

Let $n=2$, and let $L$ have a unique pole at $0$ (the simplest nontrivial case).
In this case, up to an additive constant, the operator $L$ has the form
\[
L=\partial^2+a\wp(z),
\]
where $\wp$ is the Weierstrass function of $E$ (the Lam\'e operator).
Local analysis near $0$ (i.e., the condition that the local monodromy
is trivial) shows that algebraic integrability of such $L$ implies that $a=-m(m+1)$,
where $m$ is a nonnegative integer. Conversely, it was shown by Hermite that
if $a=-m(m+1)$ then $L$ is indeed algebraically integrable.
Namely, the triviality of the monodromy of~(\ref{eige}) near~$0$
is easy to see by noting that the operator $L$ is $\Bbb Z_2$-invariant, while the indices
at~$0$ are~$-m$ and $m+1$, whose dif\/ference is odd; thus, the algebraic integrability of~$L$
follows from Theorem~\ref{fourcond}.

\subsubsection{The algebraic integrability locus}\label{section2.5.2}

Assume now that we have f\/ixed
the indices $m_j$ distinct modulo $n$ (and thus the coef\/f\/icients~$b_i$).
Then the set of possible operators $L$ is parametrized by the Laurent
coef\/f\/icients $a_{ik}$ of $a_i(z)$ of nonpositive degrees~$-k$, $k<i$, and also by the
coef\/f\/icients $g_2$, $g_3$ of the dif\/ferential equation
\[
(\wp')^2 = 4\wp^3 - g_2 \wp-g_3
\]
for the Weierstrass function $\wp$ of the elliptic curve~$E$.
Note that we have the $\Bbb C^*$-action
rescaling the lattice $\Gamma$, with respect to which
these parameters have the following degrees (or weights):
\[
\deg(g_2)=4, \qquad \deg(g_3)=6,\qquad \deg(a_{ik})=i-k.
\]
Thus, we will think of these parameters as homogeneous
coordinates on the weighted
projective space with these weights, and def\/ine
the algebraic integrability locus $X_{\bold m}$
(for each choice of indices ${\bold m}$) as a subset of
this weighted projective space.

\subsubsection{The cyclically symmetric operators}\label{section2.5.3}

For every choice of the indices ${\bold m}$, there exists a unique operator $L$
such that all the coef\/f\/icients $a_{ik}$ equal zero. Let us denote this operator by~$L(0)$.

\begin{prop}\label{cycsym}
For $n\ge 2$, the operator $L(0)$ is algebraically integrable
$($for any indices distinct modulo $n)$ in the following cases:
\begin{enumerate}\itemsep=0pt
\item[$(i)$] $n=2$  $($the Lam\'e operator any elliptic curve$)$;

\item[$(ii)$] $n=3,6$, $g_2=0$ $($equianharmonic elliptic curve$)$;

\item[$(iii)$] $n=4$, $g_3=0$ $($lemniscatic elliptic curve$)$.
\end{enumerate}
\end{prop}

\begin{proof}
In these cases, the operator $L(0)$ has a symmetry under the groups
$\Bbb Z_3$, $\Bbb Z_6$, and $\Bbb Z_4$, respectively.
This symmetry easily implies the triviality of monodromy at~$0$.
\end{proof}

\begin{rem}
Case $(i)$ is well known and due to Hermite. Case $(ii)$
was done by Halphen \cite[p.~571]{Ha}  in the case of equal gaps;
his proof easily extends to the
case of general gaps and general~$n$.
For multivariable generalizations of
Proposition~\ref{cycsym}, see~\cite{EFMV}.
\end{rem}

Note that the operators $L(0)-\lambda$, where
$L(0)$ is as in Proposition~\ref{cycsym} $(i)$--$(iii)$
are the only operators $L$ which are symmetric under $\Bbb Z_n$
(where~$n$ is the order of~$L$), acting by $z\mapsto e^{2\pi i/n}z$.
We will call such operators {\it fully cyclically symmetric}
(or just {\it cyclically symmetric} if there is no
ambiguity (e.g., when $n$ is a prime)).

\begin{conj}\label{largegaps}
For any $n\ge 2$, there exists $N\in \Bbb Z_+$ such that if $q_j\ge N$ for all $j=1,\dots ,n-1$, then
the only algebraically integrable operators $L$ with gaps $q_j$ are fully cyclically symmetric.
\end{conj}

\begin{rem}\quad {}

1.~Conjecture \ref{largegaps}, in particular,
claims that algebraic integrability for large enough gaps takes place only for
$n=2,3,4,6$.

2.~As we have explained, for $n=2$ the Conjecture \ref{largegaps} holds with $N=0$.

3.~Conjecture \ref{largegaps} is open even for $n=3$. It is supported by computational evidence
and partial results described in the next section.
\end{rem}

A similar conjecture can be made in the rational case.
Namely, assume that $a_i(z)\in \Bbb C[z^{-1}]$. In this case,
we have fully cyclically symmetric operators $L_0-\lambda$
for any $n\ge 2$, which are algebraically integrable.

\begin{conj}\label{largegaps1}
Conjecture {\rm \ref{largegaps}} holds in the rational case.
\end{conj}

\begin{rem} In the case $n=2$,
the dif\/ferential equation $L\psi=\lambda \psi$
is conf\/luent hypergeometric, and
Conjecture~\ref{largegaps1} is well known to be true
(with $N=0$).
\end{rem}

\subsection{The classif\/ication of elliptic f\/inite gap potentials}\label{section2.6}

In this subsection we give a proof of the
classif\/ication theorem of f\/inite-gap potentials
on elliptic curves with arbitrary number of poles
\cite{GW,CEO,GUW}; this proof is based on
dif\/ferential Galois theory and follows \cite[Section~4.1]{CEO}.

\begin{thm}[\cite{GUW}] \label{guw}
Let $u(z)$ be a rational function on an elliptic curve $E$,
which is a finite-gap potential $($i.e., the
operator $L=\partial^2+u$ is algebraically integrable$)$.
Then there exist
nonnegative integers $m_1,\dots ,m_N$
and points $z_1,\dots ,z_N\in E$
satisfying the equations
\[
\sum_{j\ne i}m_j(m_j+1)\wp^{(2s-1)}(z_i-z_j)=0,\qquad i=1,\dots ,N,\quad s=1,\dots ,m_i,
\]
such that
\[
u(z)=-\sum_{i=1}^N m_i(m_i+1)\wp(z-z_i)+{\rm const}.
\]
Moreover, any potential of this form is finite-gap.
\end{thm}

For the proof of this theorem, we will need
the following classical lemma
from the theory of dif\/ferential equations.

\begin{lem}\label{deq}
Let $m\in \Bbb Z_+$, and
\[
u(z)=-m(m+1)z^{-2}+\sum_{j=1}^\infty c_j z^j\in \Bbb C((z)).
\]
Then the equation
\[
(\partial^2+u)\psi=\lambda\psi
\]
admits two linearly independent solutions in $\Bbb C((z))$ for all $\lambda\in \Bbb C$
if and only if $c_1=c_3=\cdots =c_{2m-1}=0$.
\end{lem}

\begin{proof}
We need to show that the given condition is equivalent to the existence
of a solution of the form $\sum_{n\ge 0} a_n z^{n-m}$ with $a_0=1$.
If all odd-numbered coef\/f\/icients $c_j$
are zero (the $\Bbb Z_2$-symmetric case),
then the required solution clearly exists. Otherwise, the obstruction
to the existence of such solution is a certain homogeneous
polynomial $P$ of $c_j$ and $\lambda$ of degree $2m+1$ (where $\deg(c_j)=j+2$,
$\deg(\lambda)=2$). Suppose that $s$ is the smallest integer such
that $c_{2s-1}$ is nonzero. Then it is easy to compute that
the polynomial $P$ has degree $m-s$ with respect to~$\lambda$, and
its leading term is a multiple of $c_{2s-1}\lambda^{m-s}$.
Thus, this polynomial is nonzero, and the required solution does not exist.
On the other hand, if $s\ge m+1$, then $P=0$, and the required solution
exists.
\end{proof}

\begin{proof}[Proof of Theorem \ref{guw}.]
Theorem \ref{regsin} implies that
the poles of $u$ must be exactly second order (it is clear
that if $u$ has a f\/irst order
pole then (\ref{eige}) does not admit a basis of meromorphic solutions). Moreover, it follows
from Corollary \ref{intcon} that the singular part of~$u$ at each pole~$z_i$, $i=1,\dots ,N$,
is $-m_i(m_i+1)(z-z_i)^{-2}$, where $m_i$ is a nonnegative integer.
Thus, we f\/ind that in the algebraically integrable case
\[
u(z)=-\sum_{i=1}^N m_i(m_i+1)\wp(z-z_i)+{\rm const}.
\]

Let us now show that the conditions for algebraic integrability
in terms of $m_i$, $z_i$ are exactly as stated in the theorem. According to Theorem~\ref{fourcond},
the condition for algebraic integrability is that
the monodromy of~(\ref{eige}) around each pole of $u$ is trivial.
So the theorem follows from Lemma~\ref{deq}.
\end{proof}

In particular, if all $m_i=1$, we obtain the following well known
result:

\begin{cor}[\cite{AMM}]\label{calog}
The potential $u=-2\sum \wp(z-z_i)$ is algebraically
integrable if and only if $(z_1,\dots ,z_N)$ is a critical point of the
elliptic Calogero--Moser potential
\[
U(z_1,\dots ,z_N)=\sum_{1\le j\ne
i\le N}\wp(z_i-z_j).
\]
\end{cor}

\begin{rem}
The same method can be used to rederive the classif\/ication
from \cite{GW} of trigonometric and rational f\/inite-gap potentials
which are bounded at inf\/inity, i.e., potentials on
the nodal and the cuspidal curve of arithmetic genus~1.\footnote{As explained in \cite{GUW}, boundedness at inf\/inity is
in fact automatic.} It leads to the same answer, with $\wp(z)$ replaced by
$\frac{1}{\sin^2 z}$ and $\frac{1}{z^2}$, respectively.

Note that (as explained in \cite{GW} and references therein),
in the rational case we have the identity
\begin{gather}\label{summ}
\sum_i m_i(m_i+1)=m(m+1)
\end{gather}
for some $m\in \Bbb Z_+$.
This identity comes from the fact that a rational potential
\[
u(z)=-\sum_i \frac{m_i(m_i+1)}{(z-z_i)^2}
\]
can be degenerated into
\[
u_0(z)=-\frac{\sum_i m_i(m_i+1)}{z^2}
\]
by rescaling $z$ (so that algebraic integrability of $\partial^2+u$ implies algebraic
integrability of $\partial^2+u_0$).
We see that, surprisingly, identity~(\ref{summ}) follows automatically
from the existence of solutions of the equations
\[
\sum_{j\ne i} \frac{m_j(m_j+1)}{(z_i-z_j)^{2s+1}}=0,\qquad
i=1,\dots ,N,\quad s=1,\dots , m_i.
\]
\end{rem}

\section{Third order operators with one pole}\label{section3}

\subsection{The general setup}\label{section3.1}

Let us now consider in detail the case $n=3$ with one pole at
$0$. Up to an additive constant, the operator $L$ in this case looks like
\begin{gather}\label{third}
L=\partial^3+(a\wp(z)+c)\partial+(b\wp'(z)+e\wp(z)),
\end{gather}
where $a,b,c,e\in \Bbb C$.

As explained above, a necessary
condition for algebraic integrability is that
the indices $m_0<m_1<m_2$ at $0$ are integers pairwise distinct modulo $3$.
So we have $m_0+m_1+m_2=3$, and
\[
a=m_0m_1+m_0m_2+m_1m_2-2,\qquad b=\frac{1}{2}m_1m_2m_3.
\]
As above, it is convenient to introduce the ``gaps'' $q=q_1=m_1-m_0$, $r=q_2=m_2-m_1$
(then $m_0=1-(2q+r)/3$, $m_1=1+(q-r)/3$, $m_2=1+(2r+q)/3$).

Our goal is to determine the algebraic integrability locus
for each set $m_0$, $m_1$, $m_2$ in terms of the homogeneous coordinates $c$, $e$, $g_2$, $g_3$.

Recall that the $j$-invariant of $E$ is def\/ined by the formula
\[
j(E) = \frac{1728 g_2^3}{g_2^3-27 g_3^2}.
\]
In particular, the equianharmonic (i.e., $\Bbb Z_3$-symmetric) elliptic curve $E$ has $g_2=0$
and $j=0$. Thus, by Theorem~\ref{cycsym}, the point $c=e=g_2=0$ belongs to the algebraic integrability locus
for any~$q$,~$r$.

Note that the parameters $a$, $b$ of the operator $-L^*$, where $L^*$ is the adjoint operator
to $L$, are given by the formulas $a'=a$, $b'=-b+a$ (and also $c'=c$, $e'=-e$). So the indices of~$L^*$
are $2-m_i$, $i=0,1,2$, and hence the gaps $q$ and $r$
are interchanged under passing to the adjoint operator. On the
other hand, it is clear that~$L^*$ is algebraically integrable
if and only if so is~$L$. So it suf\/f\/ices to consider the case~$q\ge r$.

Also note that the gaps cannot be
divisible by $3$, and must be equal modulo $3$.
So we can write $q=r+3k$, where $k\ge 0$ is an
integer.

We have a basis of solutions of equation
(\ref{eige}) of the form
\[
\psi_i(z)=z^{m_i}(1+o(1)), \qquad z\to 0.
\]
Obviously, $\psi_2$ is single-valued near $0$;
algebraic integrability of $L$ is equivalent to the condition
that $\psi_0(z)$, $\psi_1(z)$ are
single-valued near $0$, i.e., do not contain log factors.
However, it turns out that even a weaker
condition suf\/f\/ices. Namely, we have the following important
proposition.

\begin{prop}\label{rank1}
If $\psi_1$ is single-valued near $0$ for all $\lambda$, then so is $\psi_0$, and
thus $L$ is algebraically integrable.
\end{prop}

The proof of Proposition~\ref{rank1} is based on the following
well known lemma from linear algebra.

\begin{lem}\label{abmat}
If $A$, $B$ are two square matrices such that $AB-BA$ has rank
at most~$1$ then $A$,~$B$~are simultaneously upper triangular in some
basis.
\end{lem}

\begin{proof}
Without loss of generality, we can assume $\ker A\neq 0$ (by replacing $A$ with $A-\lambda$ if needed)
and that $A\neq 0$. It suf\/f\/ices to show that there exists a proper nonzero subspace invariant
under~$A$,~$B$; then the statement will follow by induction in dimension.

Let $C=[A,B]$ and suppose ${\rm rank}\, C=1$ (since the case ${\rm rank}\, C=0$
is trivial). If $\ker A\subset \ker C$, then $\ker A$ is $B$-invariant:
if $Av=0$ then $ABv=BAv+Cv=0$. Thus $\ker A$ is the required subspace.
If $\ker A\nsubseteq \ker C$,
then there exists a vector
$v$ such that $Av=0$ but $Cv\neq 0$.
So $ABv=Cv\neq 0$. Thus $\mathrm{Im}\, C\subset \mathrm{Im}\, A$.
So $\mathrm{Im}\, A$ is $B$-invariant:
$BAv=ABv+Cv\in {\rm Im}\,A$. So ${\rm Im}\,A$ is the required subspace.

This proves the lemma.
\end{proof}

\begin{cor}\label{abmat1}
If $A$, $B$ are two invertible square matrices such that
$ABA^{-1}B^{-1}-1$ has rank
at most $1$ then $A$, $B$ are simultaneously upper triangular in some
basis.
\end{cor}

\begin{proof} Let $ABA^{-1}B^{-1}-1=T$, where $T$ has rank at most $1$.
Then $AB-BA=TBA$, and $TBA$ has rank at most $1$. So the result
follows from the lemma.
\end{proof}

\begin{proof}[Proof of Proposition \ref{rank1}.] Let $A$ and $B$ be the
monodromy matrices of (\ref{eige}) (for some base point) along
the global cycles of the elliptic curve.
Then $ABA^{-1}B^{-1}=U$, where $U$ is unipotent. Since
$\psi_1$ is single valued, $U-1$ has rank $1$. So it follows from
Corollary \ref{abmat1} that the monodromy of~(\ref{eige}) is upper
triangular, so the result follows from Theorem \ref{fourcond}.
\end{proof}

Using Proposition \ref{rank1} and CAS ``Maple'', we computed
the algebraic integrability locus for small values of $r$
and any value of $q$. In each case, we applied $L-\lambda$ to a generic function
\[
z^{m_1}\big(1+f_1z+f_2z^2+\cdots+f_r z^r+O\big(z^{r+1}\big)\big),
\]
solved for $f_1,\dots,f_{r-1}$, and look at the coef\/f\/icient of $z^{m_1+r}$
in the result.  This is a polynomial in~$\lambda$, and by Proposition \ref{rank1},
the monodromy around $0$ is trivial if\/f the coef\/f\/icients of this polynomial
vanish; this gives equations in $c$, $e$, $g_2$, $g_3$.

\begin{rem} As explained above, homothety of the elliptic curve rescales $c$, $e$, $g_2$, $g_3$ and
$\lambda$, giving a natural notion of degree:
\[
\deg(e)=1,\qquad  \deg(c)=2,\qquad \deg(g_2)=4,\qquad \deg(g_3)=6,\qquad \deg(\lambda)=3,
\]
and the constraint polynomial is homogeneous of degree $r$.
\end{rem}

The results of our computations are presented in the next subsection.

\subsection{Results of computer calculations}\label{section3.2}

\subsubsection[$r=1$]{$\boldsymbol{r=1}$}\label{section3.2.1}

The coef\/f\/icient of $z^{m_1+1}$ in the image of $L-\lambda$ is $e$, and thus
the operator is algebraically integrable if\/f $e=0$; this gives a
1-parameter family of integrable operators on each elliptic curve.
The case $q=1$ is trivial ($L=\partial^3+c\partial$), while the case $q=4$ was considered by Picard
in 1881.

\subsubsection[$r=2$]{$\boldsymbol{r=2}$}\label{section3.2.2}

The coef\/f\/icient of $z^{m_1+2}$, after solving for $f_1$, is independent of
$\lambda$, and thus gives an equation relating $c$ and $e$:
\[
c = -\frac{3e^2}{(q+1)^2}.
\]
Each curve thus has a 1-parameter family of integrable operators of this form.

In particular, setting $q=r$, $e=0$ we get Example~1 of~\cite{U}.
This is the third order operator commuting with the Lame operator
$\partial^2-2\wp$.

\subsubsection[$r=4$]{$\boldsymbol{r=4}$}\label{section3.2.3}

Now the coef\/f\/icient is linear in $\lambda$.  The leading coef\/f\/icient is a
positive multiple of~$e$, so $e=0$; then the constant coef\/f\/icient relates~$g_2$ and~$c$:
\[
c^2 = \frac{(q+2)^2}{3}g_2.
\]
Thus each elliptic curve admits a pair of integrable operators of this
form, except that when $g_2=0$, the two operators coincide
(with the corresponding cyclically symmetric operator).

In particular, setting $q=r$, $e=0$ we get Example~2 of~\cite{U}.

\subsubsection[$r=5$]{$\boldsymbol{r=5}$}\label{section3.2.4}

Again the constraining polynomial is linear in $\lambda$. The leading
coef\/f\/icient implies
\[
c = -\frac{3 (7q^2+35q+46)}{16(q+1)^2(q+4)^2}e^2
\]
at which point the constant coef\/f\/icient factors, so that either $c=e=0$ or
\[
g_2 = \frac{27(4q^2+20q+25)}{64(q+1)^4(q+4)^4} e^4.
\]
Each elliptic curve thus admits four integrable operators of this form
(or two modulo the symmetry $z\to -z$) except the equianharmonic case $g_2=0$, where
all these operators coincide with the cyclically symmetric operator.

In particular, setting $q=r$, $e=0$ we get Example 3 of~\cite{U}.

\subsubsection[$r=7$]{$\boldsymbol{r=7}$}\label{section3.2.5}

The constraint polynomial has degree 2, with leading coef\/f\/icient
\[
\frac{5(q+2)(q+5)}{144(q+1)(q+3)(q+4)(q+6)}e,
\]
so that $e$ must be 0.  But this also eliminates the constant term, leaving
only an equation relating~$c$ and~$g_2$:
\[
c^2 = \frac{25(q+2)^2(q+5)^2}{12(2q+7)^2}g_2.
\]
So we get two operators on each elliptic curve
which coincide in the $\Bbb Z_3$-symmetric case $g_2=0$ (with the cyclically
symmetric operator).

In particular, setting $q=r$, $e=0$ we get Example 4 of \cite{U}.

\subsubsection[$r=8$]{$\boldsymbol{r=8}$}\label{section3.2.6}

The constraint polynomial again has degree 2, and the leading coef\/f\/icient
implies
\[
c =
-\frac{3(191q^4+3056q^3+17598q^2+42992q+38384)}{686(q+1)^2(q+4)^2(q+7)^2}e^2.
\]
Unlike in the previous case, however, one still has two conditions
remaining, of weighted degrees~5 and 8 respectively.  The degree 5
condition naturally has a factor of $e$, but setting $e=0$ makes~$c=0$ so
that the remaining constraint is a multiple of $g_2^2$; the coef\/f\/icient is
negative, so this makes $g_2=0$.  In the remaining case, the degree~5
condition gives a formula for~$g_2$:
\[
g_2 = \frac{27g_2^{\rm num}}{g_2^{\rm den}}e^4,
\]
where
\begin{gather*}
g_2^{\rm num}:=155383 q^8+4972256 q^7+68978821 q^6+541706360 q^5+2632855228
  q^4
\\
\phantom{g_2^{\rm num}:=}{}
+8104425920 q^3+15416669104 q^2+16555419008 q+767835508
\end{gather*}
and
\[
g_2^{\rm den}=470596 (q+1)^4(q+4)^4(q+7)^4 \big(19q^2+152q+277\big),
\]
and the degree 8 condition can then be solved for $g_3/e^6$.  For each
$q=8+3k$, $k\ge 0$, one thus has a single $j$-invariant other than $j=0$
for which there exists an algebraically integrable operator:
\[
j=-6912\frac{j^{\rm num}}{j^{\rm den}},
\]
where
\[
j^{\rm num}=p_8(q)^3
(19q^2+152q+277),
\]
with\footnote{We note that $p_8$ is the numerator of $g_2/e^4$.}
\begin{gather*}
p_8(q)=155383q^8+4972256q^7+68978821q^6+541706360q^5+2632855228q^4\\
\phantom{p_8(q)=}{} +
8104425920q^3+15416669104q^2+16555419008q+7678355008
\end{gather*}
and
\begin{gather*}
j^{\rm den}=(q+7)(q+6)(q+2)(q+1)\big(67q^2+533q+898\big)\big(67q^2+539q+922\big)
\\
\phantom{j^{\rm den}=}{}
\times\big(37q^3+399q^2+
1344q+1468\big)\big(37q^3+489q^2+2064q+2692\big)\\
\phantom{j^{\rm den}=}{}
\times
\big(367q^3+5115q^2+23376q+34828)(367q^3+3693q^2+12000q+12724\big)
\\
\phantom{j^{\rm den}=}{}
\times\big(829q^4+14194q^3+89097q^2+242068q+239236\big)\\
\phantom{j^{\rm den}=}{}
\times
\big(829q^4+12334q^3+66777q^2+156028q+133156\big).
\end{gather*}
Note that $j$ is f\/inite for any $q$ since the factors of $j^{\rm den}$ have
positive coef\/f\/icients.

In particular, for $e=0$ the only solution is $g_2=0$,
which is shown for $q=r$ in Example~5 of~\cite{U}.

\subsubsection[$r=10$]{$\boldsymbol{r=10}$}\label{section3.2.7}

The constraint polynomial is cubic in $\lambda$, and the leading
coef\/f\/icient implies that $e=0$; setting $e=0$ makes the polynomial even in~$\lambda$, so one has two remaining constraints, of degrees~$4$ and~$10$
respectively.  The degree $4$ constraint can be solved for $g_2$:
\[
g_2 =
\frac{3(2069q^4+41380q^3+301017q^2+941170q+1071464)}{4400(q+2)^2(q+5)^2(q+8)^2}
c^2,
\]
at which point the degree $10$ constraint is $c^2$ times an equation for
$g_3$.  Thus either $e=c=g_2=0$ or
\[
g_3 = -\frac{g_3^{\rm num}}{g_3^{\rm den}},
\]
where
\begin{gather*}
g_3^{\rm num}:=96577q^6+2897310q^5+35259207q^4+222299140q^3
\\
\phantom{g_3^{\rm num}:=}{}
+764656215q^2+1360455150q+978817201
\end{gather*}
and
\[
g_3^{\rm den}:=
422400(q+2)^3(q+5)^2(q+8)^3.
\]
Thus other than $j=0$, the only possible $j$ invariant is
\[
j =
-995328\frac{j^{\rm num}}{j^{\rm den}},
\]
where
\[
j^{\rm num}:=\big(2069q^4+41380q^3+301017q^2+941170q+1071464\big)^3
\]
and
\begin{gather*}
j^{\rm den}=
(5q+19)(5q+31)(13q+47)(13q+83)(17q+73)(17q+97)(19q+59)\\
\phantom{j^{\rm den}=}{} \times
  (19q+131)\big(11q^2+110q+239\big)\big(23q^2+200q+317\big)\big(23q^2+260q+617\big).
\end{gather*}

\subsubsection[$r=11$]{$\boldsymbol{r=11}$}\label{section3.2.8}

The constraint polynomial is cubic in $\lambda$, and other than
$c=e=g_2=0$, there is no solution to the resulting four equations; one can
solve the f\/irst three for $c$, $g_2$, $g_3$ in terms of~$e$, and plug in to
the fourth equation, obtaining $e^{11}$ times a rational function which is
negative for~$q>8$.  Thus $c=e=g_2=0$ is the only solution.

\subsubsection[$r=13$]{$\boldsymbol{r=13}$}\label{section3.2.9}

The constraint polynomial is quartic in $\lambda$, but again the f\/irst
equation is $e=0$, and eliminates half of the remaining equations.  One
thus has two additional equations which can be solved to give either
$c=g_2=0$ or $g_2/c^2$, $g_3/c^3$ equal to specif\/ic rational functions of~$q$.  There is thus again a single surviving $j$ invariant, which is given by the formula
\[
j=-124416\frac{j^{\rm num}}{j^{\rm den}},
\]
where
\begin{gather*}
j^{\rm num}=(67q^2+871q+2014)^2
\big(24727q^6+964353q^5+15225009q^4+124224139q^3\\
\hphantom{j^{\rm num}=}{}
+551142996q^2+1258400208q+1155995968\big)^3
\end{gather*}
and
\begin{gather*}
j^{\rm den}=
(13q+68)(13q+101)\big(19q^2+265q+
796\big)\big(19q^2+229q+562\big)\\
\hphantom{j^{\rm den}=}{}\times
\big(83q^2+1094q+2936\big)
\big(83q^2+1064q+2741\big)\big(47q^3+924
q^2+5481q+9532\big)\\
\hphantom{j^{\rm den}=}{}\times
\big(47q^3+909q^2+5286q+8824\big)
\big(547q^3+14649q^2+127758q+
360056\big)\\
\hphantom{j^{\rm den}=}{}\times
\big(547q^3+6684q^2+24213q+26876\big)\big(11q^2+143q+332\big).
\end{gather*}

\subsubsection[$r=14$]{$\boldsymbol{r=14}$}\label{section3.2.10}

If $e=0$, then $c=g_2=0$; otherwise, one can solve the f\/irst three
equations for $c/e^2$, $g_2/e^4$, $g_3/e^6$, at which point the fourth
equation is again a negative multiple of $e^{11}$, so there is no additional
solution.  Similar arguments apply to $r=17, 20$.

\subsubsection[$r=16$]{$\boldsymbol{r=16}$}\label{section3.2.11}

The f\/irst equation makes $e=0$, so that one has three additional equations.
If $c=0$, then the next equation makes $g_2=0$; otherwise, one can solve
the f\/irst two equations for $g_2/c^2$, $g_3/c^3$, at which point the
remaining equation is a negative (for $q>13$) multiple of $c^8$, so no
other solution exists.  Similar arguments apply to $r=19, 22$.

\begin{rem}
Observe that the polynomials $j^{\rm den}$ in the cases $r=8,10,13$
split into many irreducible factors over $\Bbb Q$,
whose leading coef\/f\/icients are either~$1$ or primes.
Moreover, the constant coef\/f\/icients of the factors
for $r=10$ are also primes, while for $r=8, 13$ they are primes times a small
(at most third) power of~$2$. The number-theoretic
roots of this peculiar behavior are mysterious
to us. We burden the reader with the unwieldy expressions of
the $j$-invariants in the hope that someone would help us
demystify it.
\end{rem}

\begin{rem} The operators with $q=r$ for $r=10$ and $r=13$
exist only for special values of~$j$, so they are not present
in~\cite{U}, which deals with the case of generic $j$ only.
\end{rem}

\subsection{The classif\/ication conjecture}\label{section3.3}

On the basis of this data we make the following conjecture.

\begin{conj} For $q\ge r\ge 14$, there are no
algebraically integrable operators $L$ apart from the one which
is $\Bbb Z_3$-symmetric $($i.e. $c=e=g_2=0)$.
Thus, all the algebraically integrable third order operators $L$
with one pole are the ones described in this subsection.
\end{conj}

This is a more precise version of Conjecture \ref{largegaps}
for $n=3$, claiming that in this case one may take $N=14$.

Here is a partial result in the direction of this conjecture.

\begin{prop}\label{ec0}\qquad

\begin{enumerate}\itemsep=0pt
\item[$(i)$] If $r=3s+1$, where $s\ge 0$ is an integer, then
for any algebraically integrable operator~$L$
of the form \eqref{third}, one has $e=0$.

\item[$(ii)$] Assume that $r=3s+1$, where $s\ge 1$ is an integer.
If $L$ is an algebraically integrable operator of the form
\eqref{third} then $g_2=0$ if and only if~$c=0$
$($in which case~$L$ is cyclically symmetric$)$.
\end{enumerate}
\end{prop}

\begin{proof}
$(i)$ We have already checked the case $s=0$ directly, so we may assume that $s>0$.

Algebraic integrability of the operator (\ref{third}) is equivalent to the
existence of three linearly independent solutions of the dif\/ferential equation
\[
\partial^3f+\big(t^2 a\wp(tz)+t^2 c\big)\partial f + \big(t^3 b\wp'(tz) + t^3 e
\wp(tz)\big)f
=f
\]
in $\Bbb C((z))$ for generic $t$; this is just the image of the original eigenvalue equation
under a homothety $z\mapsto tz$ of scale $t=\lambda^{-1/3}$.  For $t=0$, this has three
independent solutions, each of which must deform to a solution for general
$t$ with the same asymptotics at $z=0$.  Let $\psi_1(z)$ be the middle
solution,
\[
\psi_1(z) = z^{m_1}\sum_{k\ge 0}
\frac{(z/3)^{3k}}{(1-r/3)_k(1+q/3)_k k!}.
\]
Then we need to be able to deform this to a solution of the form
$\psi_1(z)+tG_1(z)+O(t^2)$, and we claim that this implies $e=0$.  Plugging
\[
G_1(z) = \sum_{k\ge 0} c(k) z^{m_1+3k+1}
\]
into the equation gives
\begin{gather*}
\sum_{k\ge 0} [(3k+1)(3k+1-r)(3k+q+1)c(k)-c(k-1)] z^{m_1+3k-2}
\\
\qquad{}=
-e\sum_{k\ge 0}
\frac{z^{m_1+3k-2}}
     {3^{3k}(1-r/3)_k(1+q/3)_k k!}
(3k+1)(3k+1-r)(3k+q+1)c(k)-c(k-1)
\\
\qquad{}
=-\frac{e}{3^{3k}(1-r/3)_k(1+q/3)_k k!}.
\end{gather*}
Setting $c'(k)=c(k)3^{3k}(1-r/3)_k(1+q/3)_k k!$, we obtain
\[
(3k+1)(r-3k-1)(q+3k+1)c'(k)-
3k(r-3k)(3k+q)c'(k-1)=e.
\]
It follows that $c'(k)$ is a positive multiple of $e$ for
$0\le k<(r-1)/3$.  Since the equation for $k=(r-1)/3$ reads
$-(r-1)(q+r-1)c'((r-4)/3) = e$,
this gives a contradiction unless $e=0$.

$(ii)$ We have already computed the case $s=1$ directly, so we may assume
$s>1$. Also, it follows from part $(i)$ that $e=0$.  The perturbed solution,
if it exists, will thus have the form
\[
\psi_1(z) + t^2 G_1(z) + t^4 G_2(z) + O\big(t^5\big).
\]
(Note that the odd degree terms vanish by symmetry.)  To the same
order, the dif\/ferential equation reads
\[
\partial^3 f + \left(\frac{a}{z^2}+c t^2 + \frac{ag_2z^2}{20}t^4\right)\partial f
+ \left(-\frac{2b}{z^3} + \frac{b g_2 z}{10} t^4\right) f = f + O\big(t^5\big).
\]
Writing
\begin{gather*}
G_1(z)  = z^{m_1+2} \sum_k c'(k) \frac{(z/3)^{3k}}{(1-r/3)_k(1+q/3)_k k!},\\
G_2(z)  = z^{m_1+1} \sum_k d'(k) \frac{(z/3)^{3k}}{(1-r/3)_k(1+q/3)_k k!}
\end{gather*}
and substituting in, we f\/ind from the $t^2$ term that
\begin{gather*}
(3k+2)(r-3k-2)(q+3k+2)c'(k) - 3k(r-3k)(q+3k)c'(k-1)\\
\qquad{}
=
((q-r)/3+3k+1)c
\end{gather*}
and thus $c'(k)$ is a positive multiple of $c$ for $0\le k\le (r-4)/3$.
We also f\/ind that
\begin{gather*}
(3k+1)(r-3k-1)(q+3k+1)d'(k) -3k(r-3k)(q+3k)d'(k-1)
\\
\qquad{} =
-k(r-3k)(q+3k)(q-r+9k)c'(k-1)c
+
k(r-3k)(q+3k)C(k,q,r) g_2
\end{gather*}
where
$C(k,q,r)$ is positive when $4\le 3k+1\le r\le q$, since the appropriate
linear change of variables gives a polynomial with positive coef\/f\/icients.
We f\/ind by induction that $d'(k)$ is a~nonnegative linear combination of
$-c^2$ and $g_2$ for $0\le k\le (r-4)/3$, while the equation for
$k=(r-1)/3$ also tells us that $d'((r-4)/3)$ is a negative linear
combination of~$-c^2$ and~$g_2$.  Subtracting the expressions for
$d'((r-4)/3)$ gives a positive linear combination of $-c^2$ and $g_2$ which
vanishes.  In particular, if one of $c$ and $g_2$ is~0, so is the other.
\end{proof}

\subsection{The nodal and cuspidal cases}\label{section3.4}

The results of the previous subsections also apply to the nodal case
$j=\infty$ and the cuspidal case $g_2=g_3=0$.

Namely, in the nodal case,
we get 1-parameter families of algebraically integrable operators
for $r=1,2$, f\/inite collections operators for $r=4,5,7$,
and conjecturally no solutions for larger~$r$ (this is conf\/irmed
for $r\le 22$).

In the cuspidal case $g_2=0$, $g_3=0$,
we always have an algebraically integrable operator
with cyclic symmetry; apart from that, we get $1$-parameter families
of algebraically integrable ope\-rators for $r=1,2$, and
conjecturally no other cases (if $r=3s+1$, this is true by Proposition~\ref{ec0}).

\section{Operators with several poles}\label{section4}

\subsection{Third order operators}\label{section4.1}

Consider now a third order algebraically integrable operator
\begin{gather}\label{ope1}
L=\partial^3+a(z)\partial+b(z)
\end{gather}
on an elliptic curve $E$ with several poles $z_1,\dots ,z_N\in E$.

It is easy to show that if the gaps of $L$ at a given point are
$q=r=1$, then the operator must be holomorphic at this point,
i.e. this case is trivial. So we consider the simplest
nontrivial case, when the gaps at all the poles are
$q=r=2$.

\begin{lem}\label{van}
Let $L=\partial^3+a(z)\partial+b(z)$
be a Fuchsian differential operator near $z=0$
with gaps $q=r=2$. Let
\[
a(z)=\sum_{k\ge 0}a_kz^{k-2},
\qquad
b(z)=\sum_{k\ge 0}b_kz^{k-3}.
\]
Then $L-\lambda$ has trivial monodromy around $0$
for any $\lambda$ if and only if the Laurent coefficients~$a_1$,~$b_2$ are zero, and
\[
a_2=-\frac{b_1^2}{3},\qquad b_4=a_4+\frac{b_1a_3}{3}.
\]
\end{lem}

\begin{proof}
We have three gaps $2$, $2$, $4$, and three conditions associated to
them, which are of degrees $2$, $2$, $4$. The conditions of degree~$2$
say that $b_2=0$ and $a_2=-b_1^2/3$. The condition of degree~$4$
is linear in $\lambda$. The leading coef\/f\/icient in~$\lambda$
is of degree $1$, and is a nonzero multiple of~$a_1$, so we get the
condition $a_1=0$. The constant coef\/f\/icient is of degree $4$ and
gives $b_4=a_4+\frac{b_1a_3}{3}$.
\end{proof}

Clearly, the same result holds for a Fuchsian dif\/ferential
operator def\/ined near any point $z=z_0$.
Thus, any algebraically integrable operator~(\ref{ope1}) with gaps $q=r=2$ at all poles would
necessarily have to be of the form
\[
\partial^3+\left(c-3\sum_{i=1}^N \wp(z-z_i)\right)\partial
-\frac{3}{2}\sum_{i=1}^N \wp'(z-z_i)
+\sum_{i=1}^N 3p_i\wp(z-z_i),
\]
where $p_i$ and $c$ are complex numbers (up to adding a
constant). Let us f\/ind the conditions on the parameters
$z_i$, $p_i$, $c$ for this operator to be algebraically integrable.

\begin{prop}\label{cubi}
The conditions for algebraic integrability of $L$
are
\[
c+3p_i^2=3\sum_{j\ne i}\wp(z_i-z_j),\qquad i=1,\dots ,N,
\]
and
\[
\sum_{j\ne i}(p_i+p_j)\wp'(z_i-z_j)=0, \qquad i=1,\dots ,N.
\]
\end{prop}

\begin{proof}
The proof is by direct calculation using Lemma~\ref{van}.
\end{proof}

\begin{cor}\label{cub}
Let
\[
F(z,p)=\sum_{i=1}^Np_i^3-\frac{3}{2}\sum_{i\ne
j}(p_i+p_j)\wp(z_i-z_j).
\]
Then $L$ is algebraically integrable if and only
if $(z,p)$ is a critical point of
the function $F(z,p)+c\sum_{i=1}^N p_i$.
\end{cor}

We note that $F$ is the cubic integral $H_3$ for the
elliptic Calogero--Moser Hamiltonian
\[
H_2=\sum_{i=1}^N p_i^2-\sum_{j\ne i}\wp(z_i-z_j).
\]
Thus, the algebraically integrable operators for a f\/ixed value of $c$ are
the critical points of $H_3+cH_1$, where $H_1=\sum p_i$.

So, Proposition \ref{cubi}
can be viewed as a third order analog of Corollary~\ref{calog}.

\begin{rem}
Corollary \ref{cub} is, essentially, a special case of the elliptic analog of
Proposition 6 of \cite[p.~124]{AMM}.
\end{rem}

\subsection{Higher order operators}\label{section4.2}

We expect that in a similar way one can deal with higher order
operators, obtaining families of algebraically integrable
operators parametrized by critical points of higher
Calogero--Moser Hamiltonians. Specif\/ically, we expect that
if we take the $n$-th order operator $L$ with
indices $-1,1,\dots ,n-2,n$ and poles $z_1,\dots ,z_N$ on
an elliptic curve, then the algebraically integrable operators
will correspond to critical points of a degree $n$ elliptic
Calogero--Moser Hamiltonian. Similarly to Corollary~\ref{cub},
this should be a consequence of the
methods of~\cite{AMM} and~\cite{Kr2}.

This result obviously has trigonometric and rational counterparts.

We note, however, that whether the corresponding
variety of critical points is nonempty, what is its dimension,
etc., are, in general, dif\/f\/icult
questions.

\subsection{Operators with symmetries}\label{section4.3}

It is also interesting to consider operators with symmetries.
For example, suppose $L$ is a second order operator
$\partial^2+u(z)$ which is even with respect to~$z$.
Assume that it has poles at the f\/ixed points of $z\to -z$
(i.e.\ $w_0=0$, $w_1=1/2$, $w_2=\tau/2$, and $w_3=(1+\tau)/2$) and at some other
distinct points $\pm z_1,\dots ,\pm z_N$. Assume that the indices of
$L$ at the f\/ixed points~$w_i$ are $-m_i$, $m_i+1$ for $i=0,1,2,3$,
and the indices at $\pm z_j$ are $-1$, $2$. In this case, similarly
to Proposition~\ref{calog}, it is easy to show that algebraically
integrable operators correspond to critical points of the
Inozemtsev potential
\[
U:=\sum_{i=0}^3\sum_{j=1}^N
\left(m_i+\frac{1}{2}\right)^2\wp(z_j-w_i)+\sum_{1\le k\ne j\le
N}(\wp(z_j-z_k)+
\wp(z_j+z_k))
\]
(see \cite[Theorem 0.2]{Tr} for the case $N=1$).

In the same vein, one may consider operators $L$ of order
$\ell=3,4,6$ which are invariant under the group $\Bbb Z_\ell$
on an elliptic curve with such $\Bbb Z_\ell$-symmetry.
Suppose that $L$ has poles at the f\/ixed points
$\eta_j$ of $\Bbb Z_\ell$, and also at some other
points $z_1,\dots ,z_N$ (taken from distinct $\Bbb Z_\ell$-orbits)
as well as their images under the $\Bbb Z_\ell$ action.
Let us f\/ix the indices at $\eta_j$ to be the same as the indices of
the operator $L_{0j}^{\ell/\ell_j}$, where
$\ell_j$ is the order of the stabilizer of $\eta_j$,
and $L_{0j}$ is a rational homogeneous operator~(\ref{ratope}) of order $\ell_j$ with arbitrary integer indices.
Also, let us f\/ix the indices at the other poles
to be $-1,1,\dots ,\ell-2,\ell$.

\begin{conj} \label{346} Algebraically integrable
operators $L$ as above correspond to
critical points of the lowest degree $($i.e., degree $\ell)$
Hamiltonian of the classical crystallographic elliptic Calogero--Moser system
for the group $\Bbb Z_\ell$ $($with appropriate parameters$)$ defined
in~{\rm \cite{EFMV}}.
\end{conj}

\begin{rem} Conjecture \ref{346} may be generalized to the case when the indices of $L_{0j}$
are not assumed to be integers. Namely, in this case we conjecture that operators $L$ with trivial mo\-nodromy of $L\psi=\lambda \psi$ around non-f\/ixed
points of $\Bbb Z_\ell$ (i.e.\ those for which the monodromy gives rise to a representation of a generalized DAHA of rank~1
of type $E_6$, $E_7$, $E_8$ def\/ined in~\cite{EOR}) correspond to critical points  of the lowest degree
Hamiltonian of the classical crystallographic elliptic Calogero--Moser system
for the group $\Bbb Z_\ell$ with generic parameters.
\end{rem}

\begin{prop} Conjecture~{\rm \ref{346}} holds for $\ell=3$.
\end{prop}

\begin{proof}
In the case $\ell=3$, the classical crystallographic elliptic Calogero--Moser
Hamiltonian of~\cite{EFMV} has the form
\begin{gather*}
H =  \sum_{i=1}^N p_i^3 +\sum_{i=1}^N
\sum_{r=0}^2(A_r\wp(z_i-\eta_r)p_i+B_r\wp'(z_i-\eta_r))
  -3C\sum_{i\ne j}\sum_{s=0}^2 \wp(z_i-\varepsilon^s
z_j)p_i,
\end{gather*}
where $\tau=\varepsilon:=e^{2\pi {\rm i}/3}$, $\wp(x):=\wp(x,\tau)$, $\eta_0=0,\eta_1={\rm i}\sqrt{3}/3$,
$\eta_2=-{\rm i}\sqrt{3}/3$, and $A_l$, $B_l$, $C$ are parameters.

On the other hand, consider the $\Bbb Z_3$-symmetric operator
\begin{gather*}
L=\partial^3+\sum_{r=0}^2 (\alpha_r\wp(z-\eta_r)\partial+\beta_r\wp'(z-\eta_r))
-3\sum_{i=1}^N\sum_{s=0}^2\wp(z-\varepsilon^s z_i)\partial
\\
\phantom{L=}{}+
\sum_{i=1}^N\sum_{s=0}^2\left(-\frac{3}{2}\wp'(z-\varepsilon^s z_i)+3p_i\varepsilon^{-s}\wp(z-\varepsilon^s z_i)\right).
\end{gather*}
Using Lemma \ref{van} and the identity
\[
\wp\big(\big(1-\varepsilon^{\pm 1}\big)z\big)=-\frac{\varepsilon^{\mp 1}}{3}
(\wp(z)+\wp(z-\eta_1)+\wp(z-\eta_2)),
\]
we obtain the following conditions for the operator
$L$ to be algebraically integrable:
\begin{gather*}
3p_i^2=3\sum_{j\ne i}\sum_{s=0}^2\wp(z_i-\varepsilon^s
z_j)-\sum_{r=0}^2
(\alpha_r-1)\wp(z_i-\eta_r),
\\
\sum_{r=0}^2\left((\alpha_r-1)\wp'(z_i-\eta_r)p_i+\left(\frac{1}{2}\alpha_r-\beta_r\right)
\wp''(z_i-\eta_r)\right)\\
\qquad{} =3\sum_{j\ne i}\sum_{s=0}^2
\wp'(z_i-\varepsilon^s z_j)(p_i+\varepsilon^{-s}p_j),
\end{gather*}
for $i=1,\dots ,N$.
But these are exactly the conditions for
a critical point of $H$, with $A_r=\alpha_r-1$,
$B_r=\frac{1}{2}\alpha_r-\beta_r$, and $C=1$.
\end{proof}

\begin{rem} Similarly to the previous subsection,
we expect that by considering opera\-tors~$L$ of order
$n\ell$, $n>1$, with $\Bbb Z_\ell$ symmetry, one can obtain families of
algebraically integrable ope\-rators parametrized by critical
points of a Hamiltonian of degree $n\ell$ (in momenta)
for the crystallographic elliptic Calogero--Moser system
of~\cite{EFMV}.

It would also be interesting to interpret the complete f\/low
of this system (not only its critical points) along the lines
of~\cite{Kr2}.
\end{rem}

\begin{rem} Here is a rational version of Conjecture~\ref{346}, which
allows arbitrary $\ell$. Namely, let
\[
L=\partial^\ell+a_2(z)\partial^{\ell-2}+\dots +a_\ell(z)
\]
be a dif\/ferential operator with rational coef\/f\/icients,
which is invariant under $\Bbb Z_\ell$, such that $a_i(z)$ vanish at
inf\/inity. Let the nonzero poles of $L$ be $z_1,\dots ,z_N$
(taken from dif\/ferent $\Bbb Z_\ell$-orbits) as well as their images under the symmetry.
Suppose that~$L$ has arbitrary integer indices at $0$, and indices $-1,1,\dots ,\ell-2,\ell$ at
$z_1,\dots ,z_N$.

\begin{conj} Algebraically integrable operators $L$ with such properties
correspond to critical points of the rational Calogero--Moser Hamiltonian $($of degree~$\ell)$
for the complex reflection group $S_N\ltimes \Bbb Z_\ell^N$ $($see e.g.~{\rm \cite{EFMV})}.
\end{conj}

It follows from the above that this conjecture holds for $\ell\le 3$.

Moreover, we expect that a similar conjecture holds for operators of order
$n\ell$ with $\Bbb Z_\ell$-symmetry. Namely, in this case we should require
that the indices at $0$ are those of $L_0^n$, where $L_0$ is a
rational homogeneous operator
of order $\ell$ with integer indices, and we conjecture that algebraically integrable operators are
parametrized by critical points of the higher order rational Calogero--Moser Hamiltonian (of order $n\ell$).
\end{rem}

\subsection*{Acknowledgements} The authors are grateful to I.~Krichever,
E.~Previato, and A.~Veselov for useful discussions. The work of P.E.\ was partially
supported by the the NSF grants
DMS-0504847 and DMS-0854764. The work of E.R.\ was partially supported by
the NSF grant DMS-1001645.

\pdfbookmark[1]{References}{ref}
\LastPageEnding

\end{document}